\documentclass[]{spie}  

\usepackage{amsmath,amsfonts,amssymb}
\usepackage{graphicx}
\usepackage[colorlinks=true, allcolors=blue]{hyperref}
\usepackage{subfigure}
\usepackage{multirow}
\usepackage{array}
\usepackage{wasysym}
\usepackage{xcolor}
\usepackage{siunitx}
\usepackage{chemformula}

\title{Targeting low micro-roughness for 3D printed aluminium mirrors using a hot isostatic press}

\author[a*]{Carolyn Atkins}
\author[a]{Younes Chahid}
\author[a, b]{Gregory Lister}
\author[a, c]{Rhys Tuck}
\author[a]{Richard Kotlewski}
\author[d]{Robert M. Snell}
\author[d]{Elaine R. Livera}
\author[d]{Mariam Faour}
\author[d]{Iain Todd}
\author[d, e]{Robert Deffley}
\author[f]{James Shipley}
\author[f]{Tom Walsh}
\author[f]{Johannes G\aa rdstam}
\author[g]{Cyril Bourgenot}
\author[g]{Paul White}
\author[g]{Spencer Davies}
\author[h]{Samuel Tammas-Williams}

\affil[a]{UK Astronomy Technology Centre, Royal Observatory, Edinburgh, EH9 3HJ, UK}
\affil[b]{Dept of Mechanical, Aerospace \& Civil Engineering, University of Manchester, M13 9PL, UK}
\affil[c]{Dept of Mechanical, Materials \& Manu. Engineering, Uni. of Nottingham, NG7 2RD, UK}
\affil[d]{Dept of Materials Science and Engineering, University of Sheffield, Sheffield, S1 3JD, UK}
\affil[e]{Royce Translational Centre, Sheffield Business Park, Europa Ave, Sheffield S9 1ZA, UK}
\affil[f]{Quintus Technologies AB, Västerås, Sweden}
\affil[g]{Durham University, NETPark Research Institute, Sedgefield, TS21 3FB, UK}
\affil[h]{School of Engineering, University of Edinburgh, Edinburgh, EH9 3FB, UK}

\authorinfo{*Further author information, \\ E-mail: carolyn.atkins@stfc.ac.uk}

\begin{document} 
\maketitle

\begin{abstract}
Additive manufacturing (AM; 3D printing) in aluminium using laser powder bed fusion provides a new design space for lightweight mirror production. Printing layer-by-layer enables the use of intricate lattices for mass reduction, as well as organic shapes generated by topology optimisation, resulting in mirrors optimised for function as opposed to subtractive machining. However, porosity, a common AM defect, is present in printed aluminium and it is a result of the printing environment being either too hot or too cold, or gas entrapped bubbles within the aluminium powder. When present in an AM mirror substrates, porosity manifests as pits on the reflective surface, which increases micro-roughness and therefore scattered light. There are different strategies to reduce the impact of porosity: elimination during printing, coating the aluminium print in nickel phosphorous, or to apply a heat and pressure treatment to close the pores, commonly known as a hot isostatic press (HIP).

This paper explores the application of HIP on printed aluminium substrates intended for mirror production using single point diamond turning (SPDT). The objective of the HIP is to reduce porosity whilst targeting a small grain growth within the aluminium, which is important in allowing the SPDT to generate surfaces with low micro-roughness. For this study, three disks, 50 mm diameter by 5 mm, were printed in AlSi10Mg at \ang{0}, \ang{45}, and \ang{90} with respect to the build plate. X-ray computed tomography (XCT) was conducted before and after the HIP cycle to confirm the effectiveness of HIP to close porosity. The disks were SPDT and the micro-roughness evaluated. Mechanical testing and electron backscatter diffraction (EBSD) was used to quantify the mechanical strength and the grain size after HIP.
\end{abstract}


\section{INTRODUCTION}\label{sec:intro}
Aluminium alloys have a long heritage in the manufacture of low mass (lightweight) mirrors for astronomy~\cite{Chioetto2022} and Earth observation~\cite{terHorst2017}. There are a number of benefits for using aluminium in mirror production, for example, aluminium has a low density ($\sim$\SI{2.7}{\gram\per\cubic\cm}) in comparison to other metals commonly used in structures for telescope and instrument fabrication; aluminium is soft and easy to machine; and when an aluminium mirror is used in conjunction with aluminium mounts it allows the system to approach being athermal, where changes in temperature do not cause stress between mismatching materials. The fabrication of low mass aluminium mirrors is no longer restricted to subtractive (mill, drill \& lathe) or formative (casting) manufacturing methods, additive manufacture (AM; 3D printing) is a proven manufacturing method that can offer improved functionally through the layer-by-layer production, which increases design freedom. The two primary design benefits of AM in mirror fabrication are: the use of lattices and optimised geometries to create structures more tailored for the physical processes and environments that the mirror will encounter; and the ability to combine multiple, previously individual, components into one, for example, printing a mirror and the mount as one component, thereby eliminating interfaces and reducing the system part count. Further practical benefits include, reduced waste as only the required part is printed, and reduced time for manufacture of an intricate low mass structure. 

To date, AM aluminium mirrors have been printed exclusively using laser powder bed fusion (L-PBF); in this AM method, a laser is used to melt powdered aluminium and then fuse it to the previous layer. When the printing is complete the solid printed part is excavated from the volume of loose un-fused aluminium powder, which is then recycled in the next print. After printing, process chains for mirror production diverge depending whether the aluminium substrate is destined as the reflective surface~\cite{Atkins2017, Atkins2019a, Westsik2023, Herzog2015, Sweeney2015, Woodard17, Tan20, Tan22}, or if the substrate is coated in nickel phosphorus (NiP)~\cite{Heidler18, Hilpert2018, Hilpert2019, Atkins2019a, Fu2024, Atkins2018} if the goal is short wavelength applications and therefore, very low micro-roughness is required; however, the scope of this paper is where the AM aluminium substrate is used to create the reflective surface. Typical good quality micro-roughness values for AM aluminium are between \SIrange{4}{5}{\nm} root mean square (RMS, Sq) for single point diamond turning~\cite{Atkins2018, Atkins2019a, Westsik2023} (SPDT) and $<$\SI{4}{\nm} where additional polishing steps have been used~\cite{Sweeney2015, Woodard17}. However, poor quality surfaces are also frequent where the micro-roughness is $\geq$\SI{10}{\nm} RMS~\cite{Atkins2017, Atkins2019a, Westsik2023, Herzog2015, Fu2024}, the common causes for poor quality are porosity within the AM substrate, and inclusions which generate scratches. Porosity is a common defect in AM aluminium and it is a result of the printing environment being either too hot or too cold, or gas entrapped bubbles within the aluminium powder~\cite{Snell2022}. Porosity can be minimised, or locally eliminated, by optimising the printer parameters to influence the thermal environment and this is an active area of research~\cite{Snell2022, TAMMASWILLIAMS2016}. However, the thermal environment is not only reliant upon the printer parameters, it is also impacted by the mirror geometry, which adds a further layer of complexity within the optimisation. Approaching AM aluminium mirrors from the point of view of lowering cost and increasing accessibility, is there a commercially available heat treatment that can deliver a defect free reflective AM aluminium surface?

Hot isostatic press (HIP) is a heat treatment which is commonly applied to AM metal components to reduce porosity. HIP involves holding the component at high temperature and high pressure for a prolonged period of time, typically hours. HIP has been demonstrated to deliver pore free reflective surfaces with micro-roughness values between \SIrange{6}{8}{\nm}~\cite{Tan20, Tan22, Sun2024} (average, Sa), but HIP also leads to a degradation from the as-printed optimum, in addition, incorrect HIP settings can further degrade the micro-roughness~\cite{Atkins2023}. As HIP successfully removes the AM defects, the increase in the roughness is likely due to a reduction in the mechanical hardness of the material caused by a change in the microstructure.

The microstructure in L-PBF is formed during the transition from a liquid melt pool, when the laser melts the aluminium powder, into a solid structure and, in metals, crystallographic structures are formed in this transition. These crystal structures are comprised of an ordered, repeating, three dimensional array of atoms, and further, many metals are poly-crystalline, with regions of near-perfect crystal structure separated by boundary regions. Therefore the microstructure of L-PBF aluminium is an assortment of individual crystalline structures, commonly termed \textit{grains}. The grain growth in L-PBF for the common AM aluminium alloy AlSi10Mg is columnar due to its percentage of silicon (10\%) making it hypoeutectic\footnote{Note, high silicon content AM aluminium alloys, such as AlSi40 which is commonly paired with NiP in mirror fabrication~\cite{Heidler18, Hilpert2018,Hilpert2019}, are hypereutectic and do not exhibit the same grain growth.}, where the column extends in the build direction (\textit{z}), this results in a smaller cross section at \ang{0} (horizontal; \textit{x-y plane}), than at \ang{90} (vertical, \textit{z})~\cite{Kotadia2021}. The impact of the columnar grain growth is anisotropic properties with higher mechanical strength in the \textit{x-y plane}, as mechanical strength is linked to the grain diameter through the Hall-Petch relationship ($\sigma_{y}~\propto~\frac{1}{\sqrt{d}}$)~\cite{Kotadia2021}, where the yield strength ($\sigma_{y}$) is inversely proportional to the square root of the grain diameter ($d$). The effect on mirror fabrication is the expectation that reflective surfaces printed normal to the build direction at \ang{0} orientation would exhibit smaller grains, higher mechanical strength, and therefore, lower micro-roughness, than alternative orientations - confirmation of this expectation forms a foundation of this study.  

The aim of this paper is to investigate the dependence of print/grain orientation when using HIP to remove AM defects. The paper builds upon earlier studies in 2022~\cite{Snell2022} and 2023~\cite{Atkins2023}, where the micro-roughness of small (\SI{50}{\mm} diameter) mirror samples were evaluated against print orientation and the application of HIP. This new study utilises commercially available HIP settings and incorporates an analysis of the impact of HIP on reducing porosity, the mechanical strength, and the grain size. The goal of this paper is to identify optimum routes for mirror fabrication using commercially accessible processes through acknowledging the role that the microstructure has in optical quality. The experimental methodology is described in Section~\ref{sec:DoE} with the corresponding results presented in Section~\ref{sec:metrology}, the paper closes with a conclusion and future work in Section~\ref{sec:future}.   

\section{Method}\label{sec:DoE}

The 2024 study uses the same sample dimensions and print parameters as those from the previous studies. The HIP of the samples is undertaken by an industry partner to ensure that the first iteration of HIP parameters are commercially available. To assess the impact of HIP in removing porosity within the samples, X-ray computed tomography (XCT) is used to non-destructively evaluate the internal structure before and after HIP. SPDT and optical metrology of the resultant reflective surface after the post-HIP XCT, provides a quantitative description of how HIP has influenced reflectively. In the final measurements, the mirror samples are destructively tested to evaluate mechanical strength through indentation and samples are cut into smaller segments to allow electron backscatter diffraction (EBSD) to quantify the grain size and crystal orientation. The process chain is enumerated below.       

\begin{enumerate}
    \item Sample design \& additive manufacturing
    \item Rough machining \& pre-HIP XCT
    \item HIP \& post-HIP XCT
    \item SPDT \& micro-roughness evaluation
    \item Mechanical testing \& EBSD
\end{enumerate}

\subsection{Sample design \& additive manufacturing}
To provide continuity with the previous trial reported in 2022~\cite{Snell2022} and 2023~\cite{Atkins2023}, the sample dimensions and the printer settings were identical. The sample design is a \SI{50}{\milli\metre} diameter disk, with a height of \SI{5}{\milli\metre} and a \SI{1}{\milli\metre} chamfer applied to the edge of the optical surface. The sample has three `tabs', which are included for ease of handling and mounting. The samples were orientated at \ang{0} to the build plate surface (horizontal), \ang{45} and \ang{90} (vertical); support material was added to the \ang{45} and \ang{90} orientations to remove overhangs, as shown in Figure~\ref{fig:HIP_orientation}. The samples were printed using aluminium alloy AlSi10Mg, on an AconityLAB L-PBF printer (Aconity3D GmbH, Germany) using an aluminium base plate and in an argon atmosphere. The printer operated using with a \SI{390}{\watt} laser power, \SI{1.1}{\milli\metre\per\second} laser speed and a \SI{150}{\micro\metre} hatch spacing (the distance between neighbouring laser paths). Two sets of the three orientations were printed. 

\subsection{Rough machining \& pre-HIP XCT}
\begin{figure}[h]
    \centering
    \includegraphics[width = 0.94\textwidth]{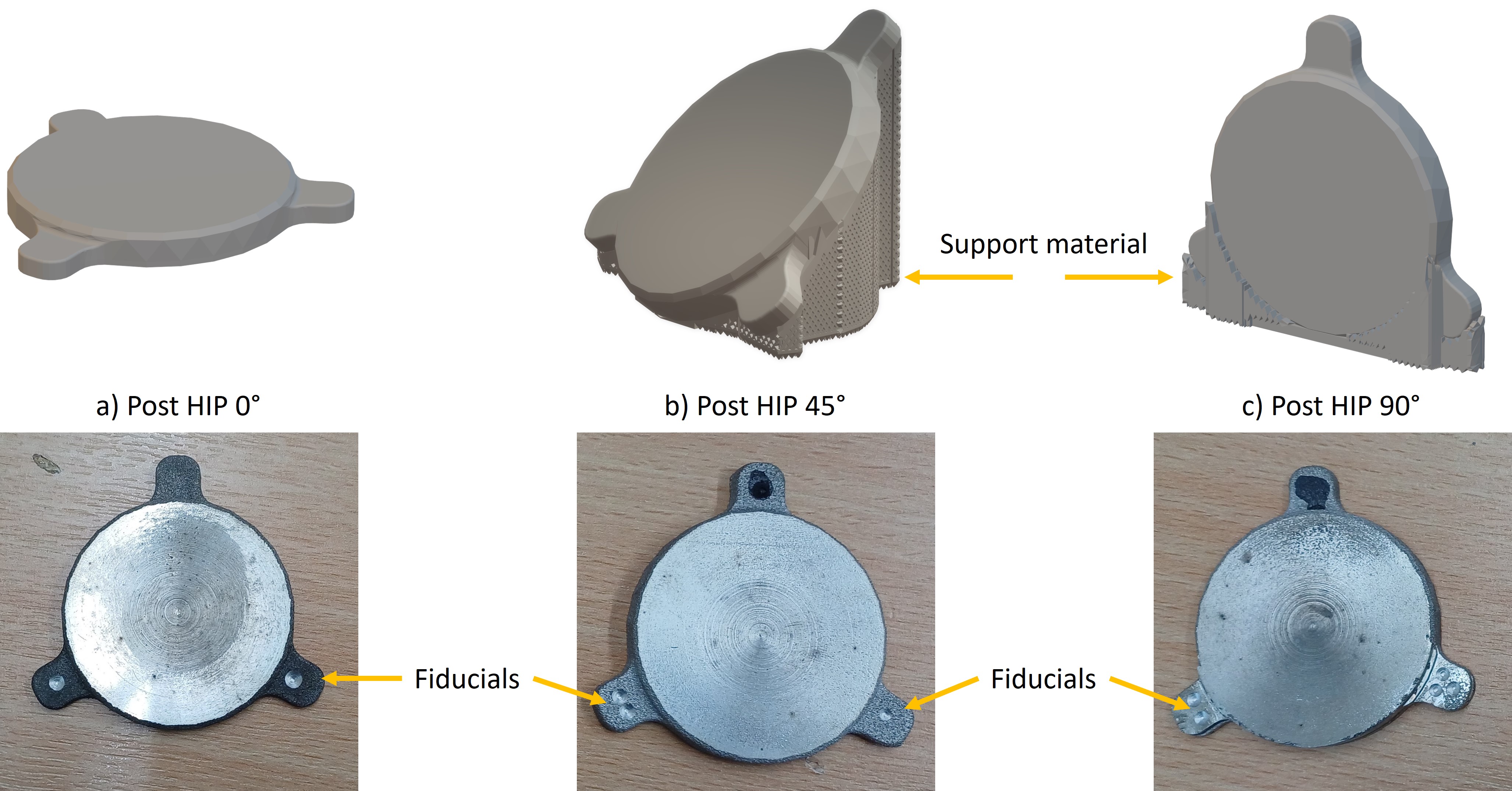}
    \caption{\textit{Upper} - the STL print files highlighting where support was required. \textit{Lower} - the post-HIP rough machined samples prior to SPDT.}
    \label{fig:HIP_orientation}
\end{figure}

Post-print, the samples were wire electrical discharge machined from the build plate and the support material removed through machining. The base and the eventual optical surface were machined to provide a smooth interface to the SPDT machine and to remove the rough post-print surface. Fiducials, drilled markers, were added to each sample to provide a unique identifier relating to its location upon the build plate and its orientation. Not only do the fiducials provide a unique identifier, they also act as alignment features when comparing the pre- and post-HIP XCT datasets. The rough machined samples are shown in Figure~\ref{fig:HIP_orientation}.

XCT measurements of each sample were taken after machining, these measurements provide a non-destructive method to assess the location, size and quantity of pores within the printed samples. During XCT, the samples are situated upon a rotating platform and positioned at a distance from the X-ray source that allows full illumination on the detector. As the platform rotates in steps, the resultant images are recorded on the detector and once the sample has rotated \ang{360}, the detector images are used to reconstruct a volumetric dataset of the sample. In this study, the samples were measured using a Nikon MCT 225 (Nikon, Japan) XCT machine with a voltage of \SI{180}{\kilo\volt}, filament current of \SI{150}{\micro\ampere}, a \SI{0.5}{\milli\metre} copper filter, and a \SI{40}{\micro\meter} voxel size. Only one volumetric dataset per sample was recorded during XCT, which limits statistical significance. 

\subsection{Hot isostatic press \& post-HIP XCT}\label{subsec:HIP}
\begin{figure}[h]
    \centering
    \includegraphics[width = 0.95\textwidth]{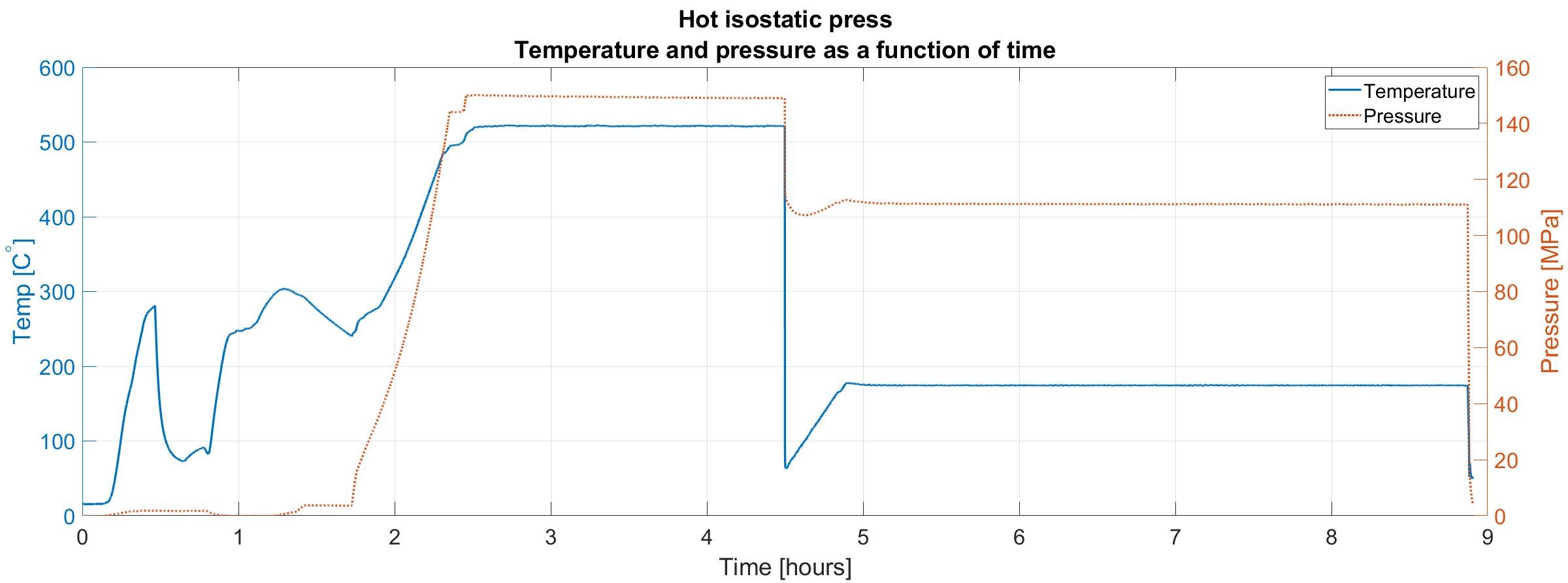}
    \caption{Temperature and pressure as a function to time for the applied HIP cycle.}
    \label{fig:HIP_plot}
\end{figure}

The 2024 samples underwent HIP with a constant maximum temperature of \SI{520}{\degreeCelsius} and a maximum pressure of \SI{150}{\mega\pascal} over a period of two hours, this was followed by a rapid quench in the HIP furnace using URQ\textsuperscript{\textregistered} (Uniform Rapid Quenching) followed by a constant \SI{175}{\degreeCelsius} and \SI{110}{\mega\pascal} over four hours. Figure~\ref{fig:HIP_plot} highlights the temperature and pressure profiles that the samples were subject to and the baseline for this study. The HIP atmosphere was argon, and clean HIP processing was used by the industrial partner, Quintus Technologies, using patented Quintus Purus\textsuperscript{\textregistered} technology. This was to ensure a clean argon atmosphere to reduce surface contamination. Note, in contrast to the Quintus HIP parameters, the 2022 \& 2023 study (RTC-HIP) used incorrect parameters, where a higher than standard temperature was recorded during the HIP ($\sim$\SI{570}{\degreeCelsius}), 

Following the HIP, XCT measurements were repeated to assess the impact of the HIP parameters. Identical scan parameters were used and the sample fiducials provided confidence comparing the pre- and post-HIP measurements. 

\subsection{Single point diamond turning \& micro-roughness}
\begin{figure}[h]
    \centering
    \includegraphics[width = 0.95\textwidth]{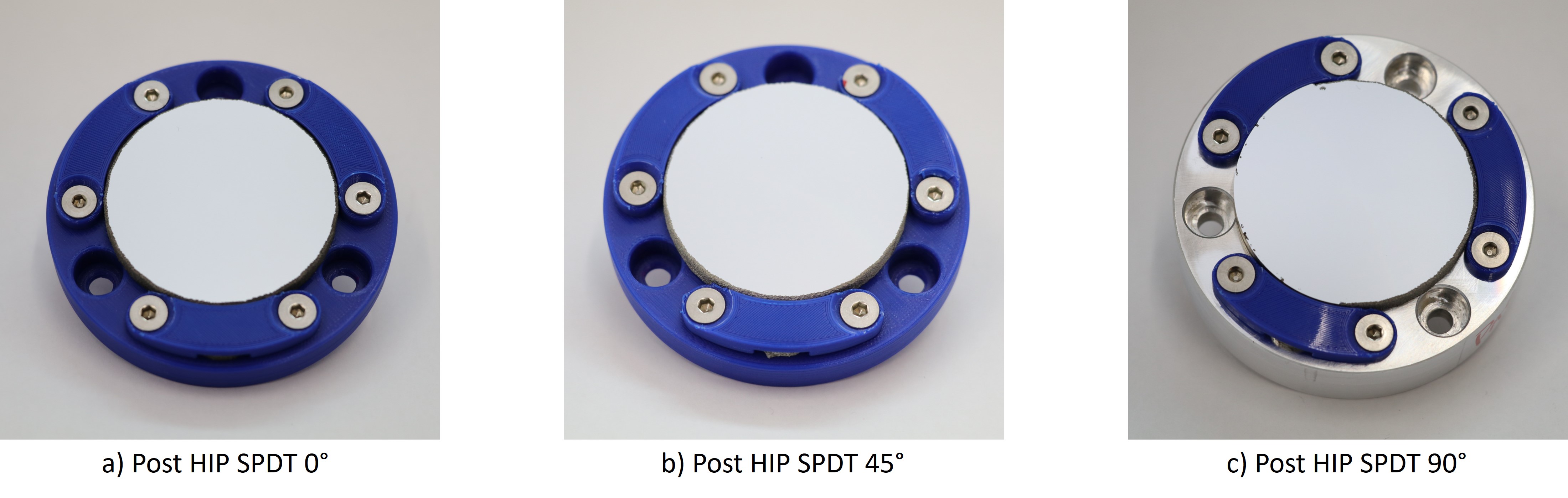}
    \caption{The reflective surface of the three orientations after SPDT.}
    \label{fig:HIP_SPDT}
\end{figure}

The samples were SPDT at the Centre for Advanced Instrumentation at Durham University using a Moore Nanotech 250 UPL precision lathe. Approximately \SI{300}{\micro\meter} was removed from the optical surface to generate the reflective surface, ensuring the removal of open pores close to the surface, between \SIrange{0}{300}{\micro\meter}. HIP is only effective at closing pores if the pore is fully encapsulated by surrounding material, pores intersecting the surface would not be closed. Figure~\ref{fig:HIP_SPDT} presents the three samples after SPDT, qualitatively all the samples demonstrated good reflectively without the presence of pits (pores) or scratches.    

An evaluation of the micro-roughness was performed using a Zygo Zegage Plus whitelight interferometer (Zygo, USA) sampling 12 locations on each surface. The measurements represented a square area of length \SI{417}{\micro\meter}, to remove low-spatial frequency distortions associated with form and waviness, an eighth order polynomial subtracted from the data.

\subsection{Mechanical testing \& electron backscatter diffraction}\label{EBSD_method}

Mechanical testing and EBSD was conducted at the University of Sheffield. Mechanical testing measured the ultimate tensile strength (UTS) of the reflective surface of the samples. The UTS was measured via profilometry-based indentation plastometry (PIP)~~\cite{Tang2021} using a PLX-Benchtop (Plastometrex, UK). In this process an indentation is made using a given load and to a defined depth, the shape of the indentation is measured using profilometry~\cite{PIP2024}. The shape of the indentation is iteratively matched using the finite element method to determine the plasticity parameters, which in turn provides the stress-strain curve to determine the yield strength and the UTS. The hardness of the material can be inferred from the UTS through the linear relationship between the two characteristics, where Vickers hardness ($H_{V}$) has an approximate relationship of three times the UTS ($\sigma_{UTS}$) ($H_{V}\approx3~\sigma_{UTS}$)~\cite{Zhang2011}. Due to the physical indentation, PIP is an optically destructive testing procedure for mirrors. 

To evaluate the size and orientation of the individual grains EBSD is used. In this measurement, incident electrons are scattered off the atoms close to the surface of the sample, if a crystal structure is present in the sample, the interactions will produce Kikuchi bands on the EBSD detector, from which information on the crystal structure and orientation is inferred. As grain boundaries typically separate regions of different crystallographic orientation, EBSD can be used to show the position of grain boundaries, and therefore provide information on the underlying grain size and morphology.

In this study, EBSD was performed in a JEOL 7900F scanning electron microscope (SEM; JEOL Ltd., Japan) with a focus depth of 21, a probe current (spot size) of 14, an accelerating voltage of \SI{20}{\kilo\volt}, a specimen tilt of \ang{70}, and step size of \SI{0.8}{\micro\meter}. Grain image and orientation analysis was performed on Matlab (Mathworks, USA) using the MTex algorithm~\cite{Bachmann2011} with a threshold orientation of $>$\ang{10} to define the high angle boundaries, and $<$\ang{3} to define the low angle grain boundaries. Grains smaller than three pixels were removed before grain reconstruction, and the resulting grains smoothed. Inverse pole figure (IPF) maps were created to represent the output from the MTex algorithm.

To extend the EBSD measurement, energy dispersive X-ray spectroscopy (EDS) was used to provide spatially resolved information on the elemental composition of a sample to investigate if any regions or features are rich or deficient in particular elements. In this case, EDS measurements are performed alongside an EBSD scan using a step size of \SI{0.18}{\micro\meter}; all other SEM settings remained constant. In this study, EDS is used clarify artefacts seen in the EBSD IPF maps. 

\section{Metrology results}\label{sec:metrology}
This section describes not only the results from the samples processed in 2024 (Quintus-HIP), but also those presented in 2023 (non-HIP \& RTC-HIP). In the case of micro-roughness (Section~\ref{subsec:micro}) previous data is drawn from 2023~\cite{Atkins2023}, whereas for mechanical testing (Section~\ref{subsec:mech}) and microstructure analysis (Section~\ref{subsec:grain}) samples from 2022~\cite{Snell2022} were tested and new data obtained. 

\subsection{X-ray computed tomography}
The analysis of the pre- and post-HIP XCT data was undertaken using Avizo 2022.1
(Thermo Fisher Scientifc, Germany). Accurate quantification of size and location of pores within an AM material is challenging due how a threshold value is determined between what is a void and what is the AM material; small variations in the threshold value can have a significant impact on both the number and size of the pores~\cite{Chahid2024}. The objective of this XCT comparison is to validate the success of the HIP to close the pores within the aluminium samples, rather than an accurate quantification of size and location. 

\begin{figure}
    \centering
    \includegraphics[width = 0.95\textwidth]{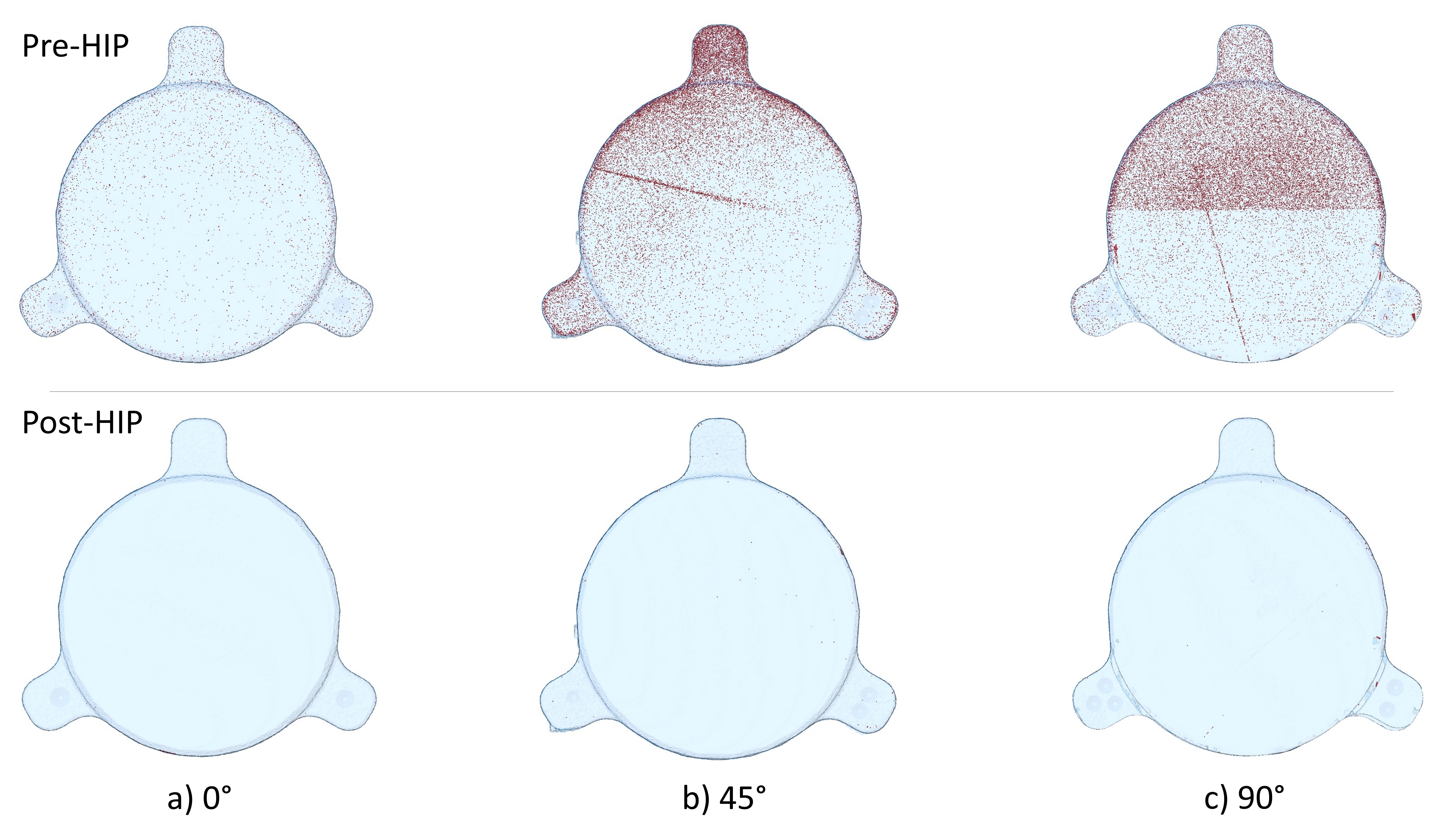}
    \caption{Porosity identification within pre- (\textit{upper row}) and post-HIP (\textit{lower row}) XCT data. Pores are identified as dark pixels within the semi transparent 3D representation of the sample.}
    \label{fig:XCT_HIP}
\end{figure}

Adaptive thresholding was used to identify the pores, this method determines local threshold values across the dataset and therefore is less affected by XCT reconstruction artefacts that locally vary the greyscale of the AM aluminium - trials with an Otsu threshold and manual thresholds could not account for a varying global greyscale. The output from the adaptive thresholding was filtered to remove `pores' that were less than five voxels, as there is low confidence in identifying pores from measurement noise at this scale. The analysis is shown in Figure~\ref{fig:XCT_HIP}, where the upper row highlights the pre-HIP data and the lower row the post-HIP data, porosity is shown as dark pixels within the structure. The individual images represent the superposition of pores within semi-transparent 3D representation. Within the figure, \ang{45} and \ang{90} have been orientated so that the top of the image correlates with the top of the sample with respect to the print orientation.

In comparing the pre- and post-HIP data, it is clear that the HIP has been successful in removing the majority of pores from the sample; however, as observed, there are features present in the pre-HIP data. A line of pixels is present in both the \ang{45} and \ang{90} pre-HIP data, these lines are seen in the original reconstructed slices and are artefacts from the reconstruction of the X-ray images into a volumetric dataset and therefore, are not porosity. However, the increase in porosity observed in the top portions of the pre-HIP \ang{45} and \ang{90}, is seen in the original slices and is considered a real effect. The \ang{90} sample provides an indication of the cause given the increase in porosity occurs between successive print layers. Between the two layers an unknown error occurred either within the printer itself, or as a result of a local change in environment. One conjecture is a change in gas flow, which impacted fume extraction and as a result, led to obscuration of the laser output and thereby limiting the power delivered to the AlSi10Mg powder. The \ang{45} pre-HIP sample does not demonstrate a clear differential between porosity levels within the orientation shown in Figure~\ref{fig:XCT_HIP}, this is in part due to the \ang{45} print orientation that creates a larger print cross section ($\frac{5~mm}{\sin{45}}$), which when viewed `face on' leads to a cross section projected with a width of \SI{5}{mm} (the disk height), this results in a linear increase in porosity density for \SI{5}{\mm} followed by a uniform density thereafter. In addition, the support material used for \ang{45} (Figure~\ref{fig:HIP_orientation} \textit{b)}), does change the thermal environment in comparison to \ang{90}, which would impact the generation of pores. Regardless, the objective of this study is the effect of HIP and, as shown, HIP has been successful at removing porosity.     

\subsection{Micro-roughness}\label{subsec:micro}
\begin{figure}[h]
    \centering
    \includegraphics[width = 0.95\textwidth]{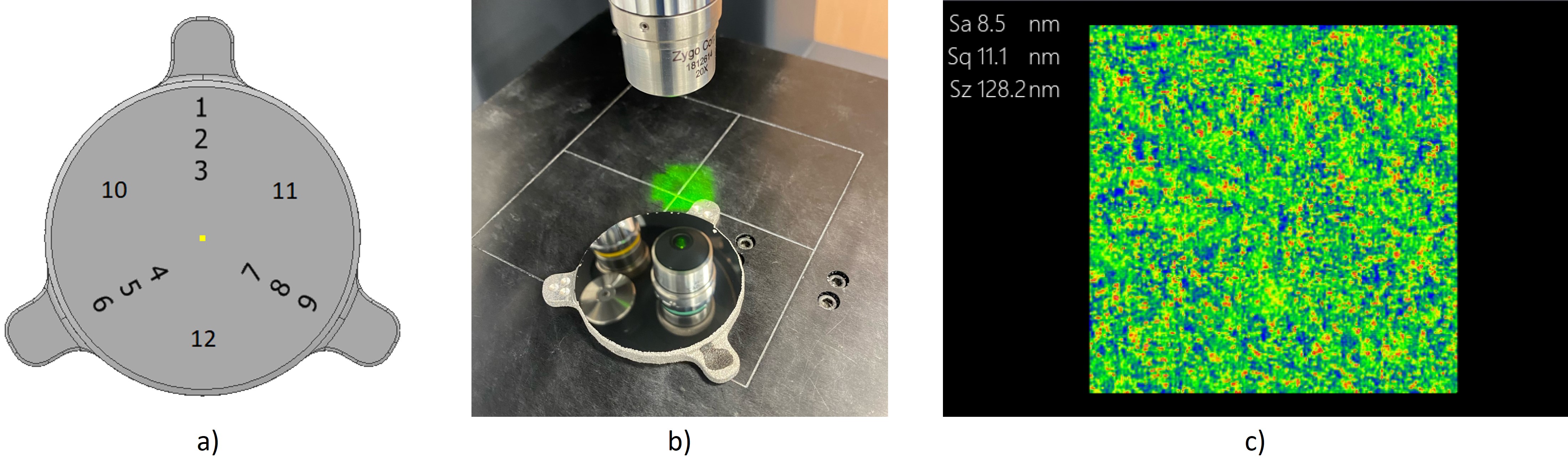}
    \caption{micro-roughness evaluation - \textit{a)} the measurement locations, \textit{b)} a sample under measurement, and \textit{c)} an representative micro-interferogram. Image credit Durham University; photo credit P. White.}
    \label{fig:met_Zygo}
\end{figure}

The micro-roughness measurement locations and output are presented in Figure~\ref{fig:met_Zygo}. In Figure~\ref{fig:met_Zygo} \textit{a)} the centre is avoided due to the SPDT residual artefact commonly produced at the centre of the spindle. Figure~\ref{fig:met_Zygo} \textit{c)} highlights a representative micro-interferogram of the optical surface. Table~\ref{tab:Zygo} collates the individual interferograms for comparison and the average, standard deviation (Std), minimum and maximum for each sample are highlighted in the final four rows. Three roughness parameters were extracted from the measurements: Sa - average roughness, Sq - root mean square roughness (RMS), and Sz - the peak to valley.

The results highlight that HIP has degraded the micro-roughness in comparison to defect-free as-printed AM AlSi10Mg~\cite{Atkins2023}; however, HIP has successfully removed the effect of porosity and scratches frequently encountered after SPDT. In comparing the three build orientations, the \ang{0} orientation provides a lower micro-roughness, Sq = \SI{9.3}{\nm} RMS, than the \ang{45} (Sq = \SI{11.5}{\nm}) and \ang{90} (Sq = \SI{11.3}{\nm}) orientations.  

{\renewcommand{\arraystretch}{1.25}
\begin{table}[h]
\centering
\caption{micro-roughness parameters extracted from the micro-interferograms for the three Quintus-HIP samples printed in orientations \ang{0}, \ang{45}, and \ang{90}.}
\begin{tabular}{|l||l|l|l||l|l|l||l|l|l|}
\hline
& \multicolumn{3}{|c||}{\ang{0} (Horizontal)}	& \multicolumn{3}{|c||}{\ang{45}} & \multicolumn{3}{|c|}{\ang{90} (Vertical)} \\
\hline
Meas.\# &	Sa [nm] & Sq [nm] & Sz [nm] & Sa [nm] & Sq [nm] & Sz [nm] & Sa [nm] & Sq [nm]& Sz [nm] \\
\hline
1 & 6.6 & 8.8 & 166.8 &  8.5 & 11.2 & 177.0 &  8.4 & 11.1 & 124.6 \\
2 & 7.1 & 9.3 & 135.0 &  8.4 & 11.1 & 126.7 &  8.5 & 11.1 & 128.2 \\
3 & 7.3 & 9.8 & 427.1 &  8.5 & 11.1 & 116.8 &  8.5 & 11.2 & 140.8 \\
4 & 7.1 & 9.4 & 160.3 &  8.7 & 11.4 & 143.9 &  8.8 & 11.5 & 180.0 \\
5 & 6.9 & 9.2 & 221.1 &  8.7 & 11.5 & 126.0 &  8.6 & 11.3 & 202.2 \\
6 & 6.7 & 9.0 & 348.3 &  9.1 & 12.2 & 264.2 &  8.7 & 11.5 & 205.3 \\
7 & 7.0 & 9.2 & 135.5 &  8.8 & 11.6 & 132.5 &  8.8 & 11.6 & 422.2 \\
8 & 6.9 & 9.0 & 163.3 &  8.7 & 11.5 & 155.7 &  8.7 & 11.3 & 158.9 \\
9 & 6.7 & 9.0 & 248.4 &  9.0 & 11.9 & 237.9 &  8.9 & 11.6 & 144.8 \\
10& 7.2 & 9.5 & 259.7 &  8.6 & 11.3 & 157.8 &  8.6 & 11.3 & 129.4 \\
11&	6.9 & 9.4 & 332.1 &  8.4 & 11.1 & 125.2 &  8.8 & 11.4 & 135.2 \\
12&	7.0 & 9.5 & 362.4 &  8.7 & 11.5 & 119.8 &  8.6 & 11.2 & 138.0 \\
\hline
\multicolumn{10}{|l|}{}\\
\hline
\textbf{Ave.} & \textbf{7.0} & \textbf{9.3} & \textbf{246.7} &  \textbf{8.7} & \textbf{11.5} & \textbf{157.0} &  \textbf{8.7} & \textbf{11.3} & \textbf{175.8} \\
Std               & 0.2 & 0.3 & 95.8 &  0.2 & 0.3 & 45.7 &  0.1 & 0.2 & 79.0 \\
min.               & 6.6 & 8.8 & 135.0 &  8.4 & 11.1 & 116.8 &  8.4 & 11.1 & 124.6 \\
max.               & 7.3 & 9.8 & 427.1 &  9.1 & 12.2 & 264.2 &  8.9 & 11.6 & 422.2 \\
\hline
\multicolumn{10}{l}{}\\
\end{tabular}
\label{tab:Zygo}
\end{table}}

\subsection{Mechanical Testing}\label{subsec:mech}
\begin{figure}[h]
    \centering
    \includegraphics[width = 0.95\textwidth]{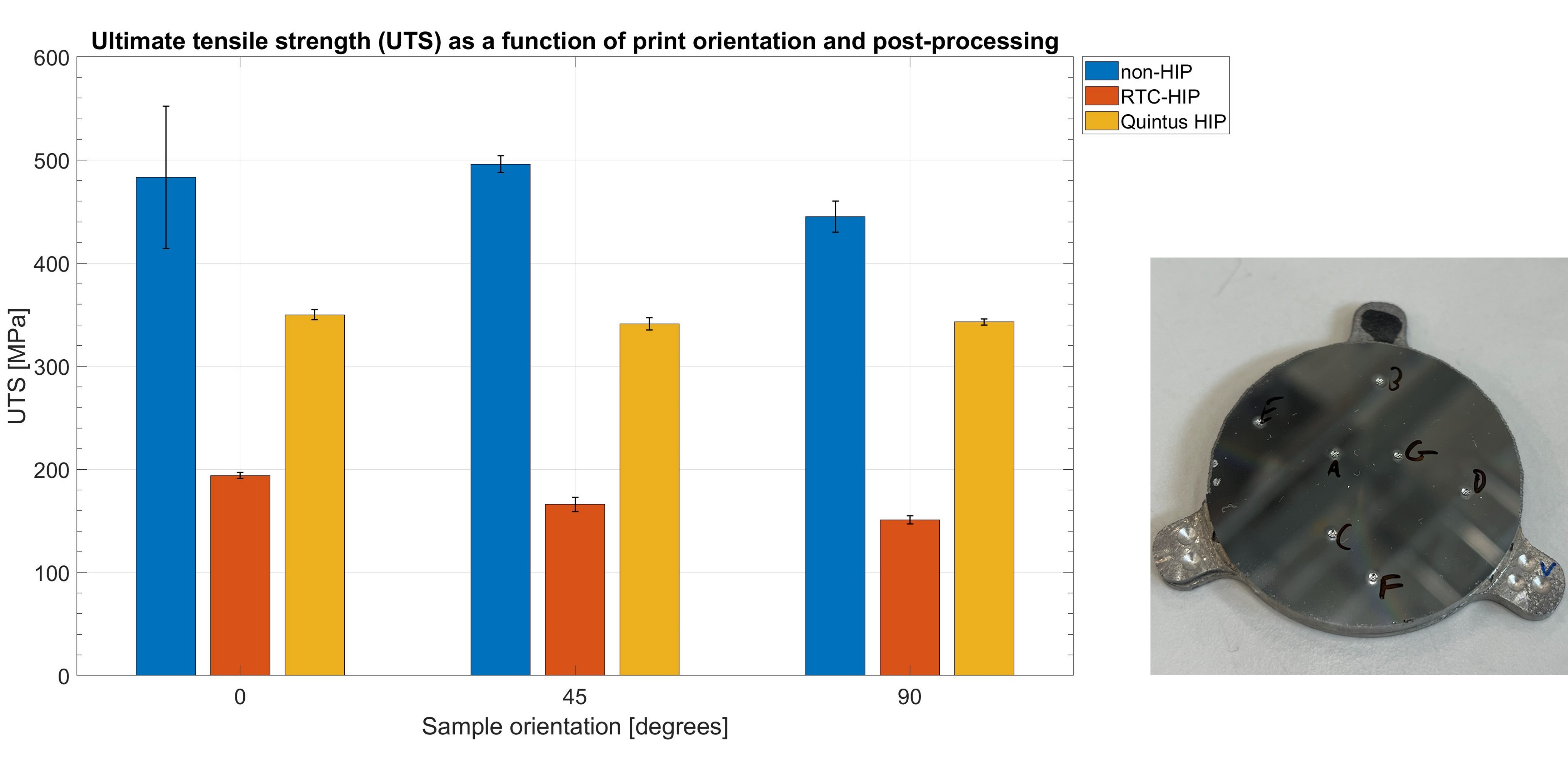}
    \caption{\textit{left} The average UTS measurements, with standard deviations, for the three post-processing variations and three build orientations, and \textit{right} a photo highlighting a sample after indentation (photo credit: R. M. Snell).}
    \label{fig:UTS}
\end{figure}

PIP was used to evaluate the UTS of three post-processed mirror variations: the non-HIP and RTC- (Royce Translational Centre) HIP mirrors presented in 2023~\cite{Atkins2023}; and the Quintus-HIP mirrors created in this study. For each variant, three print orientations were measured, resulting in 9 samples for comparison. The objective for comparing the three different post-processing methods was to link the micro-roughness with empirical hardness data. Table~\ref{tab:UTS} presents the average UTS values for the different post-processing and build orientations; the statistical values, standard deviation (std), minimum and maximum, are provided to highlight variation across the sample. The number of measurements per sample is provided in the final row. Figure~\ref{fig:UTS} \textit{left} provides a graphical comparison between the different build orientations and post-processing methods as a function of UTS, and \textit{right} a photo presents the sample after test. 

{\renewcommand{\arraystretch}{1.25}
\begin{table}[h]
\centering
\caption{The ultimate yield strength (UTS) as a function of post-processing (HIP) and build orientation}
\begin{tabular}{|l||l|l|l||l|l|l||l|l|l|}
\hline
& \multicolumn{3}{|l||}{non-HIP}	& \multicolumn{3}{|l||}{RTC-HIP} & \multicolumn{3}{|l|}{Quintus-HIP} \\
\hline
UTS [MPa] & 0$^{\circ}$ & 45$^{\circ}$ & 90$^{\circ}$ & 0$^{\circ}$ & 45$^{\circ}$ & 90$^{\circ}$ & 0$^{\circ}$ & 45$^{\circ}$ & 90$^{\circ}$ \\
\hline
Ave. & 483 & 496 & 445 & 194 & 166 & 151 & 350 & 341 & 343 \\
Std & 69 & 8 & 15 & 3 & 7 & 4 & 5 & 6 & 3 \\
Min. & 342 & 490 & 430 & 191 & 156 & 147 & 341 & 336 & 339 \\
Max. & 519 & 509 & 464 & 196 & 173 & 155 & 354 & 352 & 348 \\
\# meas. & 6 & 5 & 4 & 3 & 5 & 4 & 7 & 7 & 7 \\
\hline
\end{tabular}
\label{tab:UTS}
\end{table}}

Table~\ref{tab:UTS} and Figure~\ref{fig:UTS} demonstrate that the non-HIP mirrors have a higher UTS, and therefore hardness, than the HIP mirrors. This data is in agreement with the micro-roughness data that indicate non-HIP mirrors, ignoring the effect of porosity, exhibit higher reflectivity than HIP mirrors. HIP has the effect of reducing the UTS/hardness which leads to an increase in micro-roughness and therefore a decrease in reflectivity. Selecting HIP parameters has a clear effect on the UTS and the resultant micro-roughness. As mentioned in Section~\ref{subsec:HIP}, the RTC-HIP experienced a higher than standard HIP temperature ($\sim$\SI{570}{\degreeCelsius}), which resulted in a `milky' appearance upon the optical surface and a high micro-roughness value. It can now be shown that the UTS values for these samples have been reduced by over half the from the equivalent non-HIP values. In contrast, the Quintus-HIP, which reached a maximum temperature of $\sim$\SI{520}{\degreeCelsius}, has an approximate factor of two increase in the UTS than the RTC-HIP and this is replicated as an increase in micro-roughness within the RTC samples. 

Statistically, the standard deviation of the measurements indicate, with the exception of non-HIP \ang{0}, that the UTS is relatively constant across the samples. The origin of the anomalous data point in non-HIP \ang{0} is unknown, it could be a measurement error, but it has the effect of reducing the average UTS in that sample. Excluding the anomalous point results in an average UTS of \SI{511}{\mega\pascal} with a standard deviation of $\sim$\SI{7}{\mega\pascal}. 

Comparing between build orientations, there is evidence that the UTS is higher at \ang{0} and \ang{45} in the non-HIP, and this trend is replicated in the RTC-HIP. The Quintus-HIP has reduced the variation between orientations; \ang{0} Quintus-HIP is $\sim$\SI{10}{\mega\pascal} greater than \ang{45} and \ang{90}. However, contrasting with the micro-roughness (Table~\ref{tab:Zygo}), the \ang{0} Quintus-HIP demonstrates a clear improvement in micro-roughness (Sq $\sim$\SI{9}{\nm}) versus the \ang{45} and \ang{90} (Sq $\sim$\SI{11}{\nm}). Due to the marginal difference in UTS Quintus-HIP data, alternative data sources are required to provide further insight into the difference seen in micro-roughness between the orientations.    

\subsection{Grain size and orientation analysis}\label{subsec:grain}
EBSD scans evaluated the three \ang{0} mirrors; non-HIP, RTC-HIP and Quintus-HIP; measurements are on-going for \ang{45} and \ang{90} orientations. Samples, approximately \SI{10}{\mm} length, \SI{5}{\mm} width and \SI{5}{\mm} depth, were extracted from the middle of the mirror surfaces and polished using a standard metallurgy preparation procedure. A secondary polishing step was required to remove deformation from the samples and this was achieved using argon ion-polishing using a PECS II (Gatan, Inc.). EBSD scans were taken using the JEOL 7900F SEM and the resultant IPF maps were orientated with respect to the AM build direction (\textit{z}).

\begin{figure}[h]
    \centering
    \includegraphics[width = 0.95\textwidth]{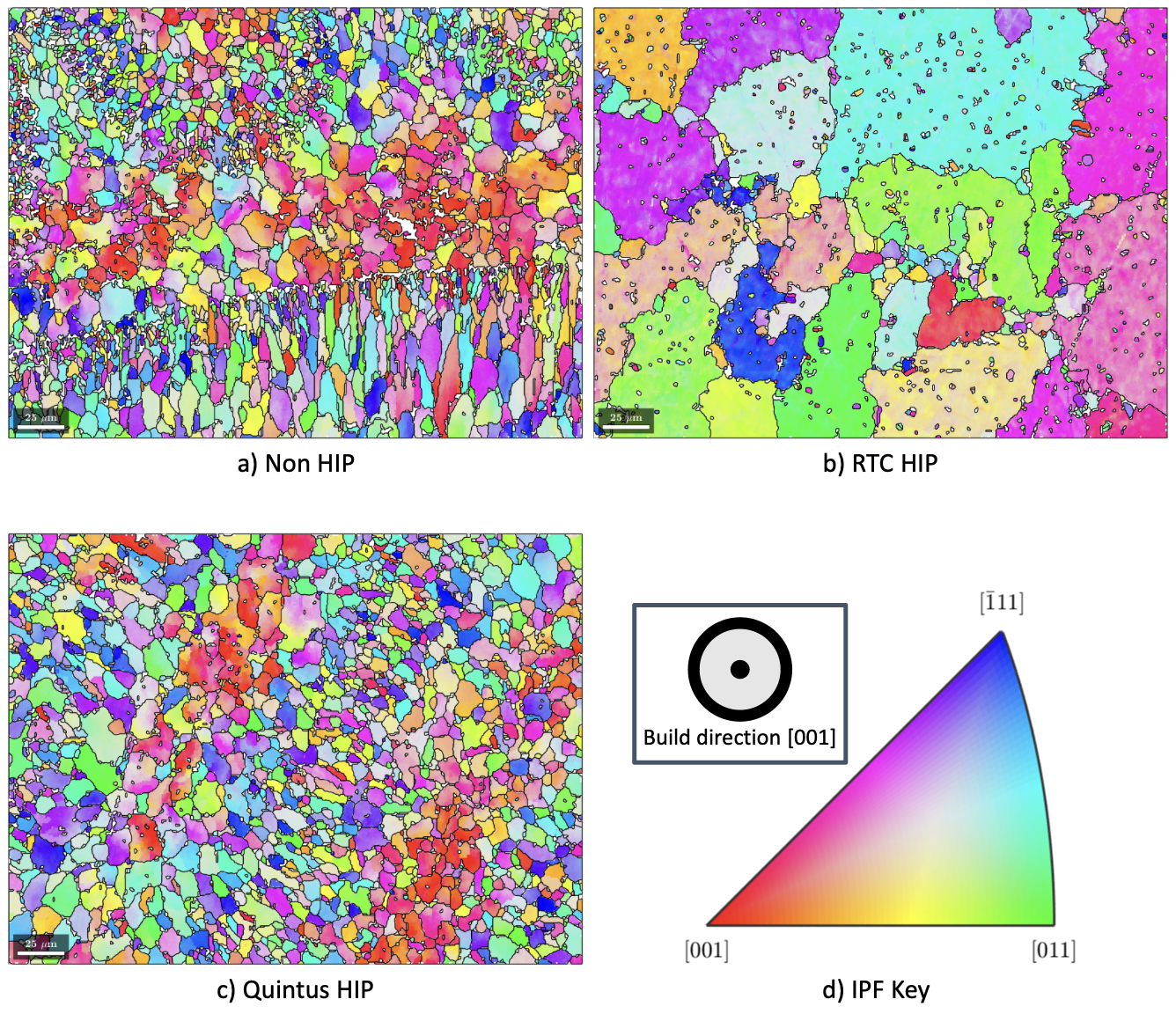}
    \caption{EBSD IPF maps with respect to the build direction for the \ang{0} mirrors: \textit{a)} non-HIP, \textit{b)} RTC-HIP, and \textit{c)} Quintus-HIP. In the bottom right, d), the inverse pole figure key is shown, showing the relationship between grain colour on the plots and the grain orientation. The orientation [001] represents the AM build direction \textit{z}.}
    \label{fig:EBSD}
\end{figure}

AM AlSi10Mg shows columnar grain growth in the build direction, the cross section of which is evidenced by a line of fine grain structure, as demonstrated in Figure~\ref{fig:EBSD} \textit{a)} of the non-HIP sample. The red coloured grains indicate that the preferential growth direction of the crystal is oriented parallel to the build direction, out of the mirror surface. These fine grains form in the middle of the laser path, where there is the maximum temperature gradient. Next to the fine grains are elongated grains that have formed at the edge of the melt pool; these grow within the plane of the mirror surface.

The RTC-HIP shows significant increase in grain size in comparison to the non-HIP, as shown in Figure~\ref{fig:EBSD} \textit{b}). The majority of grains in this map are an order of magnitude greater in cross sectional area than the non-HIP. As observed, the grain characteristics from the AM process, the columnar growth and melt pool edge grains, have been removed during the RTC-HIP. In contrast, the Quintus-HIP (Figure~\ref{fig:EBSD} \textit{c})) contains grains closer in appearance to the non-HIP sample, with evidence of columnar fine grains (red) and melt pool edges. There are fewer fine grains than in the non-HIP sample, but the grain coarsening is minimal in comparison to the RTC-HIP.  

\begin{figure}[h]
    \centering
    \includegraphics[width = 0.95\textwidth]{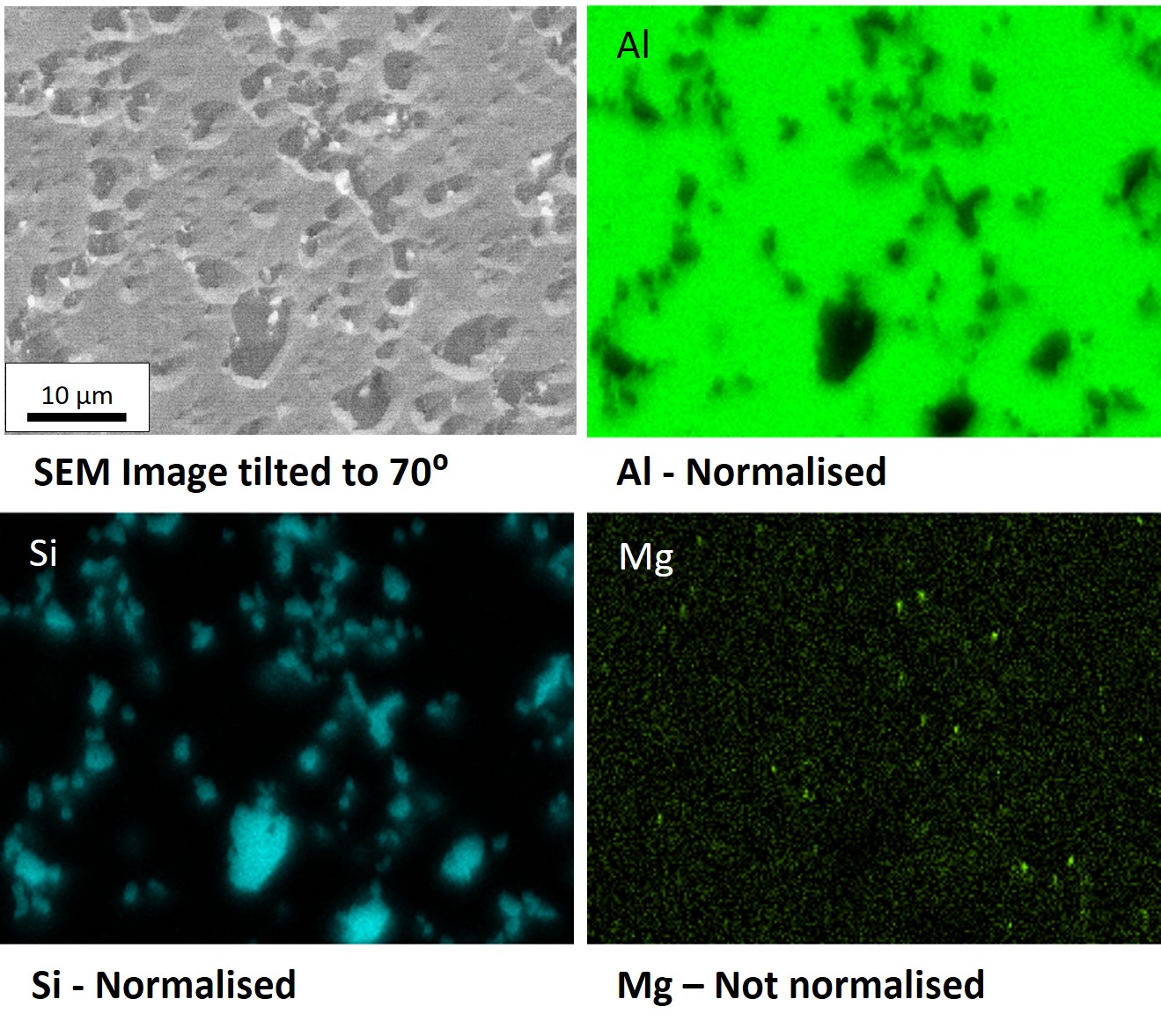}
    \caption{EDS images of the precipitates from the RTC-HIP mirror surface. The \textit{top right} and \textit{bottom} row images represent element scans, as labelled. High concentrations of elements are shown as bright colours, with lower concentrations as black. The \textit{top left} represents a standard SEM image, showing where the element distributions originated.}
    \label{fig:EDS}
\end{figure}

EDS was used to identify the elemental composition of the small granular artefacts observed within the large grains in the RTC-HIP IPF map (Figure~\ref{fig:EBSD} \textit{b}). The dominant elemental compositions are shown in Figure~\ref{fig:EDS} and it should be noted that the brightness of each image cannot be used to provide comparative information on the amount of each element present. From the EDS data and analysis using Thermo-Calc Software, there is confidence that the artefacts observed in the RTC-HIP grains are \ch{Mg2Si} or \ch{Si} precipitates mis-indexed to the aluminium phase. The precipitates are visible as dark regions in the SEM image in Figure~\ref{fig:EDS} and the majority are silicon rich and aluminium deficient; this is expected given the composition of the alloy (\ch{Al} balance \& \ch{Si} $\sim$10\%) and the elevated temperatures involved in the HIP process. Literature suggests that the EDS observed precipitates have a high likelihood of being \ch{Si} precipitates~\cite{Megahed2022, Lam2015, Zhou2018}, especially given the low magnitude of \ch{Mg} observed (Figure~\ref{fig:EDS} \textit{bottom right}).

\subsection{Discussion}\label{subsec:disc}

By using a multi-metrology approach, the impact of build orientation and HIP has been investigated towards developing a strategy for aluminium AM mirrors. XCT demonstrated the effectiveness of HIP in removing internal porosity and loose material, which led to SPDT mirrors with no pits or scratches. However, although a consistent surface quality was achieved, evidenced by the low standard deviation in the micro-roughness measurements, the roughness was higher than that exhibited in the areas with no defects within the non-HIP. Mechanical testing and microstructure analysis (EBSD \& EDS) highlighted the loss in mechanical strength (UTS) through the increase in grain size after HIP and demonstrated the importance of correct HIP parameters to minimise grain growth and to avoid precipitates.

To date, the impact of build orientation is only partially determined; however, it is hoped that future EBSD measurements will provide microstructure data consistent with the expected columnar grain growth with respect to the build orientation. The micro-roughness data demonstrated a noticeable improvement between the \ang{0} and \ang{45} \& \ang{90} orientations and, with lower confidence, this was confirmed in the UTS data. Linking microstructure with micro-roughness, would provide a useful design metric in AM mirror manufacture, where if the priority is optimum micro-roughness after HIP, then a mirror design designed for a \ang{0} orientation would be preferred. However, there is likely to be a trade-off between optimum micro-roughness and the practicalities of AM designs rules and the lightweighting requirements of the AM mirror.

\section{Conclusion \& future work}\label{sec:future}

The goal of this paper was to add fundamental knowledge towards a strategy for using AM to create low mass mirrors and, specifically, this paper adds knowledge for the wavelength applications from visible (VIS) $\longrightarrow$ infrared (IR), where micro-roughness requirements are less challenging. Aluminium continues to be a versatile material for low mass mirror fabrication and AM will allow new opportunities to enhance design for function; however, AM defects need to be considered. From the results presented, and supported via the summary of AM Al mirrors in the Introduction (Section~\ref{sec:intro}), HIP is a quick, low cost, and effective heat treatment to deliver a uniform micro-roughness across the optical surface, but due to the increase in grain size, and consequently Sq value, applications are currently limited to the near-IR (\SIrange{750}{1400}{\nm}) - assuming $<$5\% scatter is a requirement of the mirror or system. To adopt AM Al mirrors in VIS applications, alternative approaches are required to minimise the impact of AM defects in the reflective surface. 

Minimising AM defects in aluminium without HIP can be split into two approaches: AM print parameters, and post-processing. AM print parameters can be adapted for specific attributes, for example, the laser path and power can be adjusted to create an optimum print to approach zero porosity~\cite{Snell2022}; or the parameters can be adjusted to reduce porosity in a given location, for example, near the optical surface~\cite{TAMMASWILLIAMS2016}. However, although this approach has been proven on sample cubes and cylinders, a low mass mirror substrate creates a different thermal environment and an optimised setting on a cube does not necessarily translate into an optimised setting on a low mass mirror geometry. The AM print parameters approach is desirable as it removes porosity from the source, but it should be noted that commercial AM bureaux are often reluctant to adjust AM print parameters, which limits broad availability. The second approach, post-processing, accepts porosity as an attribute of AM and adds a further post-processing step. There is evidence that shot peening, or bead blasting, is an effective process in reducing porosity between \SIrange{0}{500}{\micro\meter} from the surface~\cite{Damon2018}. To implement this process effectively in mirrors, shot peening should occur after rough machining and before SPDT.  

In addition, there are two approaches that focus on minimising the microstructure: underbuild for HIP, and grain refiners. The first approach actively under-builds a sample with high porosity but fine microstructure, and then uses HIP to densify the material. The aim in this approach is to target a post-HIP microstructure comparable to a commercially printed microstructure. The final approach is the addition of grain refiners to the AlSi10Mg powder to limit grain growth. Previous research has demonstrated the potential of \ch{TiB2} and \ch{LaB6} to limit grain growth in AlSi10Mg~\cite{Kotadia2021}; however, how these grain refiners would affect SPDT and the reflective surface is unknown.     

Future work will focus on completing the microstructure analysis of the \ang{45} and \ang{90} orientations to provide clarity on the role of build orientation to affect micro-roughness. With a complete experimental data package for the Quintus-HIP, a second HIP iteration is planned with bespoke settings with the aim to limit grain growth. Proof-of-concept studies exploring the potential of shot peening to limit porosity in the vicinity of the SPDT layer are anticipated in 2024. Research is on-going to adapt the AM method specifically for mirror development, by optimising the print parameters and, in the future, the application of grain refiners. In the long-term, the research presented in this study needs to feedback into a practical application and to be applied in low mass mirrors intended for space-based astronomy and Earth observation. In this feedback, space qualification and optical requirements need to be considered, particularly in how the different approaches might impact optical performance in a thermally varying, harsh environment over time.      

\acknowledgments 
The authors acknowledge the UKRI Future Leaders Fellowship ‘Printing the future of space telescopes’ under grant \# MR/T042230/1 and the Henry Royce Institute for Advanced Materials for access the Aconity3D AconityLAB facilities at The Royce Discovery Centre at the University of Sheffield, funded through EPSRC grants EP/R00661X/1, EP/S019367/1, EP/P02470X/1 and EP/P025285/1. Collaborators from the University of Sheffield acknowledge the EPSRC Future Manufacturing Hub in Manufacture using Advanced Powder Processes (MAPP) EP/P006566/1.

\bibliography{report} 
\bibliographystyle{spiebib} 

\end{document}